\documentclass[preprint2]{aastex}

\slugcomment{Not to appear in Nonlearned J., 45.}

\shorttitle{o-p H$_2$ conversion in dense cores}
\shortauthors{Bovino et al.}

\usepackage{natbib}
\usepackage{array}
\usepackage{booktabs}
\usepackage{url}
\usepackage{lipsum}

\begin{document}


\title{H$_2$ ortho-to-para conversion on grains: A route to fast deuterium fractionation in dense cloud cores?}

\author{S. Bovino\altaffilmark{1,3}, T. Grassi\altaffilmark{2}, D. R. G. Schleicher\altaffilmark{3}, and P. Caselli\altaffilmark{4}}
\affil{$^1$Hamburger Sternwarte, Universit\"at Hamburg, Gojenbergsweg 112, 21029 Hamburg, Germany}
\affil{$^2$Centre for Star and Planet Formation, Niels Bohr Institute \& Natural History Museum of Denmark, University of Copenhagen, \O ster Voldgade 5-7, DK-1350 Copenhagen, Denmark}
\affil{$^3$Departamento de Astronom\'ia, Facultad Ciencias F\'isicas y Matem\'aticas, Universidad de Concepci\'on, Av. Esteban Iturra s/n Barrio Universitario, Casilla 160, Concepci\'on, Chile}
\affil{$^4$Max-Planck-Institut f\"ur extraterrestrische Physik, Giessenbachstrasse 1, 85748 Garching, Germany}
\email{stefano.bovino@uni-hamburg.de}






\begin{abstract}
Deuterium fractionation, i.e. the enhancement of deuterated species with respect to the non-deuterated ones, is considered to be a reliable chemical clock of star-forming regions. This process is strongly affected by the \textit{ortho}-to-\textit{para} (o-p) H$_2$ ratio. In this letter we explore the effect of the \mbox{o-p H$_2$} conversion on grains  on the deuteration timescale in fully depleted dense cores, including the most relevant uncertainties that affect this complex process. 
We show that (i) the o-p H$_2$ conversion on grains is not strongly influenced by the uncertainties on the conversion time and the sticking coefficient and (ii) that the process is controlled by the temperature and the residence time of ortho-H$_2$ on the surface, i.e. by the binding energy. We find that for binding energies in between \mbox{330-550 K}, depending on the temperature, the o-p H$_2$ conversion on grains can shorten the deuterium fractionation timescale by orders of magnitude, opening a new route to explain the large observed deuteration fraction $D_\mathrm{frac}$ in dense molecular cloud cores. 
Our results suggest that the star formation timescale,  when estimated through the timescale to reach the observed deuteration fractions, might be shorter than previously proposed. However, more accurate measurements of the binding energy are needed to better assess the overall role of this process.

\end{abstract}


\section{Introduction}
Stars form in dense and cold regions within molecular clouds, and this process is regulated by different physical phenomena which act at small scales. Magnetic  pressure, turbulence, and rotation in fact can delay the collapse inducing long timescales for stars to form. These processes are particularly relevant to determine  timescales for the formation of massive stars. 

Despite their fundamental astrophysical importance the pathways of high-mass star formation remain highly controversial \citep[see][]{Zinnecker2007}.  
In particular, do they form through slow collapse supported by magnetic pressure and/or turbulence \citep{McKee2003}, or during a fast global collapse process \citep{Bonnell2001}?  That is, the {\it timescales} of the process remain elusive to date.  Chemistry, and in particular deuteration chemistry, provides a tool to measure these timescales \citep{Fontani2011}. A high deuteration fraction, $D_\mathrm{frac}$, i.e. the ratio of deuterated to non-deuterated species, has been observed in infrared dark clouds (IRDCs).
The latter are  characterised by large amount of molecular freeze-out \citep[e.g.][]{Hernandez2011,Chen2011,Giannetti2014}, which favors a high level of deuteration \citep{Caselli2002,Fontani2011}. Several studies reported a [D/H] ratio well above the expected cosmic value of $\sim$10$^{-5}$, with values ranging from 0.001 to 0.1 \citep{Fontani2011,Chen2011,Barnes2016,Kong2016ApJ}.

Deuterium fractionation is a complex problem which depends on the interplay between many different processes and very peculiar conditions \citep{Ceccarelli2014,Kortgen2017}. It starts to be effective when low temperatures favor the freeze-out of molecules like CO on the surface of grains \citep{Tafalla2002}. Molecular freeze-out, together with the initial H$_2$ ortho-to-para ratio (OPR) regulates the speed of the deuteration process and determines its timescale. CO and \mbox{o-H$_2$} are known to destroy H$_2$D$^+$, a key molecule for deuteration,  slowing down the entire process \citep{Ceccarelli2014}. While CO depletion could be estimated from observations, the H$_2$ OPR cannot be inferred \citep{Pagani2009,Pagani2013}. H$_2$ does not emit cooling radiation in cold regions and cannot be detected unless probed indirectly through other tracers like H$_2$CO \citep{Troscompt2009}. This means that the knowledge of the H$_2$ OPR merely comes from theoretical studies. The values assessed by modelling observations vary from 0.1 in the pre-stellar core L183 \citep{Pagani2009},  10$^{-4}$ toward IRAS 16293-2422 \citep{Brunken2014}, to higher values when moving to diffuse environments \citep{Crabtree2011}. It is therefore crucial to model the H$_2$ OPR taking into account all relevant processes. In particular, the o-p H$_2$ conversion on grains so far has not been taken into account for studies related to deuteration in dense molecular cloud cores. \citet{Bron2016} have shown that this conversion process is efficient in photodissociation regions if temperature fluctuations of dust grains are taken into account. 

Recent  theoretical and experimental works have found different timescales for this process on different substrate materials.

Morphology, porosity effects, binding energy and sticking coefficient experiments, have been conducted both on bare silicates \citep[e.g.][]{Vidali2010} and porous and non-porous materials, including amorphous solid water (ASW) mixed with methanol or O$_2$ impurities \citep[e.g.][]{Chehrouri2011}. The formation of molecules on the surface has been shown to have a two-values distribution of binding energies and then different timescales \citep[e.g.][]{Hornekaer2005,Watanabe2010}. If species are adsorbed in deep sites instead of surface peaks, the diffusion is slower as well as the desorption, producing larger binding energies.

The subject of this Letter is the o-p H$_2$ conversion on dust and the impact it might have on the evolution of deuteration in dense cores. This process is affected by several uncertainties that we will discuss here.  

In the following sections we will introduce the main processes needed to study the o-p H$_2$ conversion on dust, and present the results obtained from one-zone calculations. We will then draw our conclusions.

\section{Nuclear spin conversion rate}
The conversion from ortho-to-para H$_2$ spin states in the gas phase proceeds via reactions with H$^+$ and H$_3^+$ \citep{Hugo2009,Honvault2011}.
While the above reactions are considered to be the main paths for the H$_2$ nuclear conversion, in the last years the o-p conversion on dust has attracted a lot of attention \citep{Watanabe2010,Ueta2016}. The latter depends on the interplay among different processes: i) the adsorption of o-H$_2$ on grains regulated by the adsorption coefficient $k^{\mathrm{oH_2}}_{ads}$, ii) the o-p conversion time $\tau_{conv}$, and iii) the desorption or residence time $t_{des}$. The o-p H$_2$ conversion probability (in units of s$^{-1}$) is then defined as

\begin{equation}
	P = k^{\mathrm{oH2}}_{ads} \eta\\,
\end{equation}
with $\eta$ representing the efficiency of the process \citep{Fukutani2013} given as

\begin{equation}
	\eta = \frac{t_{des}}{t_{des} + \tau_{conv}} \\.
\end{equation}
The conversion process strongly depends on the residence time of o-H$_2$ on the grain surface: if $t_{des} \gg \tau_{conv}$, the molecule will reside on the grain surface for long enough  to maximise the conversion probability (i.e. $\eta \rightarrow 1$). On the contrary, if  $t_{des} \ll \tau_{conv}$, the residence time is too short to allow efficient conversion, and $\eta \rightarrow 0$ as well as the probability.

The adsorption coefficient is

\begin{equation}
	k^{\mathrm{oH_2}}_{ads} (T_{gas}) =  S v_{\mathrm{oH_2}} \sigma_{gr}\\,
\end{equation} 
where $v_{\mathrm{oH_2}} = \sqrt{\frac{8 k_b T_{gas}}{\pi m_{\mathrm{oH_2}}}}$ is the thermal gas speed, $m_{\mathrm{oH_2}}$ the mass of o-H$_2$, $T_{gas}$ the gas temperature, $k_b$ the Boltzmann constant, $\sigma_{gr}$ the distribution-averaged grain geometrical cross-section, and $S$ the sticking coefficient.

The desorption time is
\begin{equation}\label{eq:des}
	t_{des} = \frac{1}{k_{des}} = t_0 \exp\left(\frac{E_d}{k_b T_{dust}}\right)\\,
\end{equation}
where $E_d$ is the molecule binding energy on the grain surface, $T_{dust}$ the dust temperature here assumed to be equal to $T_\mathrm{gas}$, and

\begin{equation}
	\nu_0 = \frac{1}{t_0} = \frac{1}{\pi}\sqrt{\frac{2 E_d}{d_0^2 m_{\mathrm{oH_2}}}}
\end{equation}
the harmonic frequency \citep{Hasegawa1992,Bron2016}, 
which, taking a typical width of the surface potential well of $d_0$~=~0.1~nm \citep{Bron2016}, is $\nu_0=$10$^{13}$ s$^{-1}$.

We here assume that the desorption occurs mainly through evaporation, neglecting photo-desorption and grain collisions. Photo-desorption induced by cosmic rays could be relevant at low-temperatures even in dense environments  \citep[e.g.][]{Reboussin2014}, but in the current study this turns out to be much longer than the conversion time $\tau_{conv}$.

We also neglect the inverse process which converts p-H$_2$ to o-H$_2$ as this is in general much slower, with timescales larger than \mbox{$10^4$ s} \citep[see e.g. Tab. 1 in][]{Bron2016}.

\paragraph{Conversion time} Different values for the o-p H$_2$ conversion time ($\tau_{conv}$) have been reported, and these strongly depend on the surface morphology. In particular, the presence of paramagnetic defects could induce fast spin-flip phenomena \citep{Chehrouri2011}.

\citet{Sugimoto2011} proposed a model of electric-field induced nuclear spin-flip where the electronic states of the two isomers are mixed by the Stark effect. And recently, \citet{Ueta2016} measured temperature-dependent conversion times, and found that the latter cannot be explained by the difference in the potential sites but is induced by two-phonon energy dissipation process which directly correlates with the temperature.

Overall, the range of conversion times reported in the literature lies in between $220-10^4$~s, where the higher end value comes from an extrapolation of data obtained by \citet{Watanabe2010} and then should be carefully employed. 
These values are collected in Tab.~\ref{tab:binding} for temperatures of 10 and 15~K. 

\begin{figure*}[ht]
	\centering
	\includegraphics[scale=0.42]{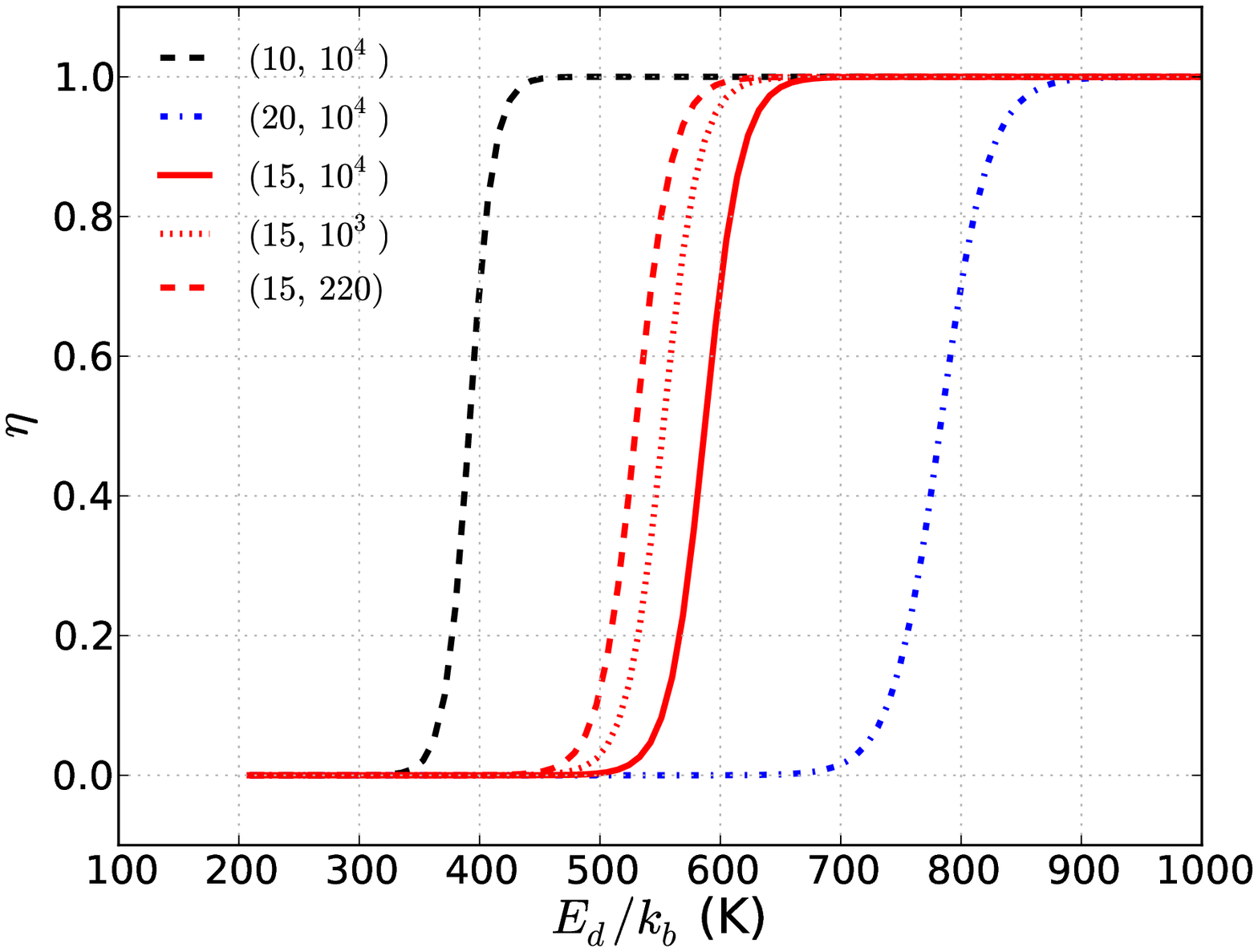}
	\includegraphics[scale=0.42]{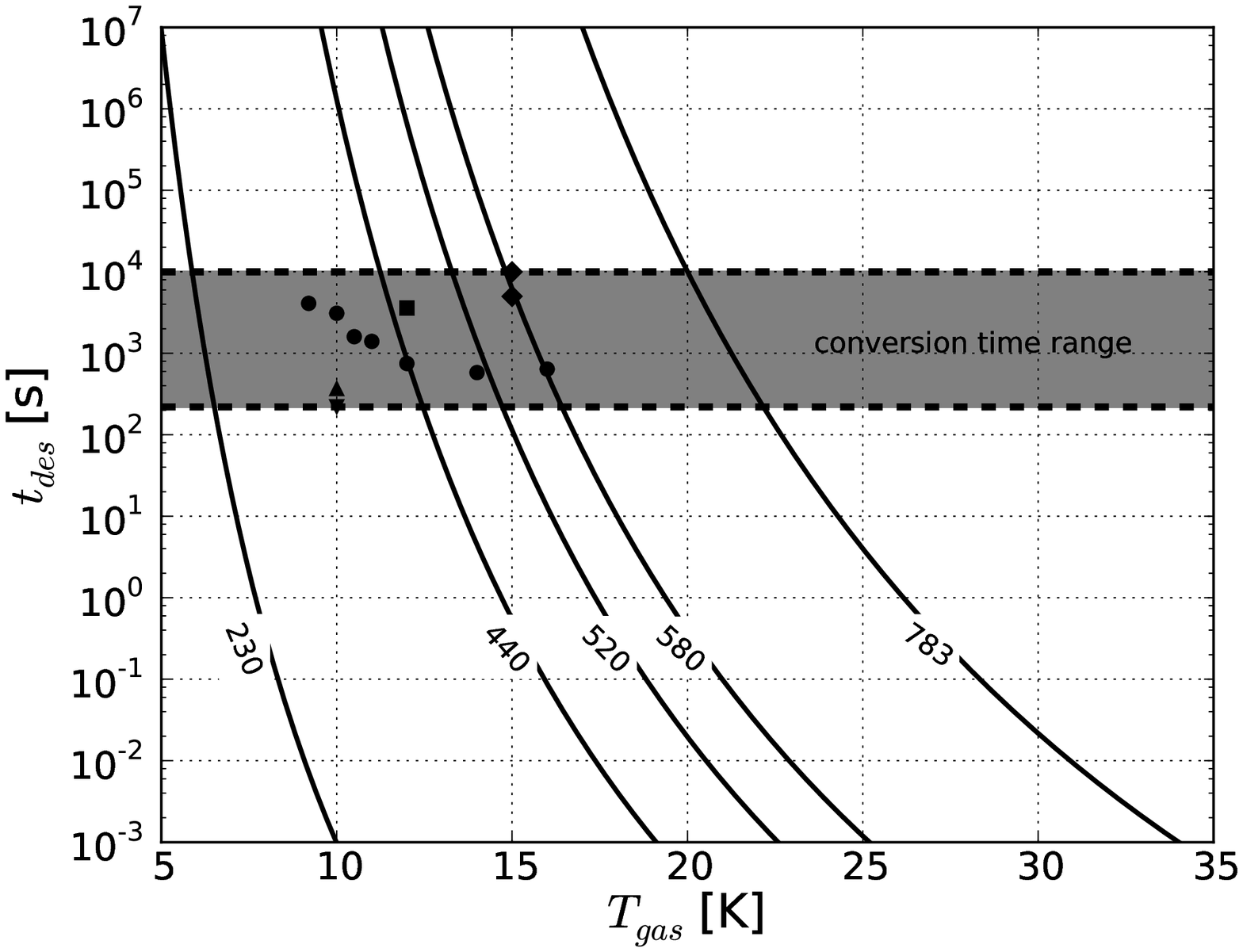}
	\caption{Left: conversion efficiency $\eta$ as a function of the binding energy $E_d/k_b$ in K, for different temperatures and conversion times $\tau_{conv}$, labelled as ($T_{gas}, \tau_{conv}$). The solid red line represents our reference run, with $T_{gas} = 15$~K and $\tau_{conv} = 10^4$~s. Right: desorption time $t_{des}$ for different binding energies (see labels), and as a function of temperature. We assume $T_{dust} = T_{gas}$. The same panel reports the o-p H$_2$ conversion times from different experiments (symbols). The grey band represents the interval of measured $\tau_{conv}$.}\label{fig:analysis1}
\end{figure*}

\begin{table*}
  \centering
  \caption{A: Binding energies for different surface types. For the two-values cases the low- and high-end values represent binding energies for the shallow and deep sites, respectively. B: Conversion times for different experiments. See text for details.
  References: (1) \citet{Cuppen2007}, (2) \citet{Amiaud2015}, (3) \citet{Watanabe2010}, (4) \citet{Al-Halabi2007}, (5) \citet{Dulieu2005}, (6) \citet{Hornekaer2005}, (7) \citet{Roser2002}, (8) \citet{Manico2001}, (9) \citet{Buch1993}, (10) \citet{Perets2007}, (11) \citet{Vidali2010}, (12) \citet{Acharyya2014},  (13) \citet{Ueta2016}, (14) \citet{Chehrouri2011}, (15) \citet{Sugimoto2011}.}\label{tab:binding}
  \begin{tabular}{*{6}{p{.16\linewidth}}}
  \toprule
    \multicolumn{6}{c}{A: Binding Energies} \\ \toprule
    \multicolumn{2}{p{.33\linewidth}}{Ref.} & \multicolumn{2}{p{.33\linewidth}}{$E_d/k_b$ (K)} & 
      \multicolumn{2}{p{.33\linewidth}}{Surface type} \\ \midrule
    \multicolumn{2}{l}{(1)} & \multicolumn{2}{l}{440} & \multicolumn{2}{l}{np-ASW} \\
     \multicolumn{2}{l}{(2)} & \multicolumn{2}{l}{783} & \multicolumn{2}{l}{p-ASW} \\
	\multicolumn{2}{l}{(3)} & \multicolumn{2}{l}{230 and 580} & \multicolumn{2}{l}{ASW} \\     
     \multicolumn{2}{l}{(4)} & \multicolumn{2}{l}{650} & \multicolumn{2}{l}{ASW} \\
     \multicolumn{2}{l}{(5)} & \multicolumn{2}{l}{520} & \multicolumn{2}{l}{np-ASW} \\
          \multicolumn{2}{l}{(6)} & \multicolumn{2}{l}{893 and 1090} & \multicolumn{2}{l}{p-ASW} \\
               \multicolumn{2}{l}{(7)} & \multicolumn{2}{l}{522 and 789} & \multicolumn{2}{l}{ASW heat-treated} \\
               \multicolumn{2}{l}{(7)} & \multicolumn{2}{l}{453 and 778} & \multicolumn{2}{l}{ASW low-density} \\
               \multicolumn{2}{l}{(7)} & \multicolumn{2}{l}{615 and 801} & \multicolumn{2}{l}{ASW high-density} \\
               \multicolumn{2}{l}{(8)} & \multicolumn{2}{l}{731} & \multicolumn{2}{l}{ASW high-density} \\
               \multicolumn{2}{l}{(9)} & \multicolumn{2}{l}{500} & \multicolumn{2}{l}{ASW} \\               
              \multicolumn{2}{l}{(10)} & \multicolumn{2}{l}{406 and 615} & \multicolumn{2}{l}{bare silicates} \\
               \multicolumn{2}{l}{(11)} & \multicolumn{2}{l}{662} & \multicolumn{2}{l}{bare silicates} \\
             \multicolumn{2}{l}{(12)} & \multicolumn{2}{l}{480} & \multicolumn{2}{l}{bare silicates} \\
     \bottomrule
    \\
    \multicolumn{6}{c}{B: Conversion times} \\ \toprule
    \multicolumn{2}{p{.33\linewidth}}{Ref.} & \multicolumn{2}{p{.33\linewidth}}{$\tau_{conv}$ (s)} & 
      \multicolumn{2}{p{.33\linewidth}}{$T_{dust}$ (K)} \\ \midrule
    \multicolumn{2}{l}{(13)} & \multicolumn{2}{l}{3.1$\pm 0.7\times$10$^3$} & \multicolumn{2}{l}{10} \\
    \multicolumn{2}{l}{(14)} & \multicolumn{2}{l}{220} & \multicolumn{2}{l}{10} \\
	\multicolumn{2}{l}{(15)} & \multicolumn{2}{l}{370$^{+340}_{-140}$} & \multicolumn{2}{l}{10} \\     
	\multicolumn{2}{l}{(13)} & \multicolumn{2}{l}{580$\pm 0.3$} & \multicolumn{2}{l}{15} \\
	\multicolumn{2}{l}{(3)} & \multicolumn{2}{l}{5$\times$10$^3$} & \multicolumn{2}{l}{15} \\
	\multicolumn{2}{l}{(3)} & \multicolumn{2}{l}{10$^4$} & \multicolumn{2}{l}{15} \\
    \bottomrule
  \end{tabular}
\end{table*}

\paragraph{Binding energy} The conversion probability depends on the desorption time (Eq.~\ref{eq:des}), i.e. on the binding energy ($E_d$). Experimental and theoretical works have been intensively dedicated to the measurements/calculations of the H$_2$ binding energy on different solid surfaces, and in function of the coverage factor $\theta$, i.e.~the number of adsorbate particles divided by the number of adsorption sites on the surface\footnote{A complete monolayer has $\theta=1$.}. Former studies focused on porous-ASW (p-ASW) with high levels of coverage, while recent experiments have been conducted on non-porous ASW (np-ASW) and as a function of the coverage \citep[e.g.][]{He2016}. Interstellar ices are indeed thought to have low levels of porosity, as they are continuously exposed to external radiation \citep{Palumbo2006}. In the next Section we will investigate the effect of the o-p H$_2$ conversion on dust assuming different binding energies\footnote{We indicate the binding energy value with the corresponding temperature $E_d/k_b$.} in the range $230-1090$ K.

\section{Chemical model}
To follow the time evolution of $D_\mathrm{frac}$ and the H$_2$ OPR, we perform one-zone calculations at constant density and temperature typical of the observed dense cores. To reduce the uncertainties and to have a better comparison with previous works, we employ the reduced fully-depleted network of \citet{Walmsley2004} updated to \citet{Hugo2009}.\footnote{As reported by \citet{Sipila2010} the new rates affect the evolution of H$_3^+$ decreasing the steady state value by a factor of three. The $D_{frac}$ is then increased by the same factor as the steady state value of H$_2$D$^+$ is not affected (see Fig.~2 in their paper).}. 
Dust grains are included using a size-distribution $\varphi(a)$, where chemical rates have been re-calculated consistently. We employ the MRN distribution $\varphi(a)\propto a^{-3.5}$ \citep{Mathis1977} in the range $a_{min} = 5\times 10^{-7}$~cm and \mbox{$a_{max} = 2.5\times 10^{-5}$~cm}, and a dust to gas ratio $\mathcal{D}$=0.013 \citep{Walmsley2004}. The grain bulk mass density is 3~g~cm$^{-3}$, typical of silicates\footnote{We tested different grain size distributions, but the effects are not relevant and we then omit the discussion.}. Given the employed density we assume \mbox{$T_{dust} = T_{gas}$.}

The chemical rate equations are integrated with the chemistry package 
 \textsc{krome}\footnote{\url{www.kromepackage.org}, commit: e2a1a54.} \citep{Grassi2014}.

\section{Results}
In the following, before discussing our results, we present a timescale analysis to understand what are the most relevant physical parameters that influence the o-p H$_2$ conversion.

In Fig.~\ref{fig:analysis1} (left) we plot the o-p conversion efficiency $\eta$ as a function of the binding energy for different $T_{gas}$ and conversion times, while Fig.~\ref{fig:analysis1} (right) reports desorption times calculated from Eq. \ref{eq:des} for different binding energies and $T_{gas}$.
We first note that, by changing the binding energy, a sharp transition of the conversion efficiency $\eta$ appears. Second, a change in temperature has a strong impact on the binding energy where the transition $\eta=0$ to \mbox{$\eta=1$} occurs, since  \mbox{$t_{des}\propto\exp(1/T_{gas})$}, i.e. increasing the temperature reduces exponentially the desorption time. Vice-versa, reducing the temperature increases the desorption timescale,  the o-H$_2$ residence time, and the conversion probability. Overall, the conversion time, which represents one of the main uncertainties together with the binding energies, does not show a strong impact on the o-p conversion efficiency while binding energy does. 

For a temperature of 15~K, when $E_d>600$~K, there is no dependence on the conversion time. On the other hand, for $E_d < 600$~K, changing $\tau_{conv}$ from $10^4$ to $220$~s, strongly affects the conversion efficiency. Fig.~\ref{fig:analysis1} (right) also indicates that the desorption time and the binding energy play a relevant role here. We have a clear indication about the temperature range where the o-p H$_2$ conversion becomes relevant once the binding energy is set. We report five different binding energies for different types of surface morphologies to explore the typical range of binding energies and material employed in laboratory experiments. 

Considering the uncertainties, we found that when $T_{gas}\leq17$~K and $E_d\simeq 580$~K, the o-p H$_2$ conversion on grains is already efficient because the conversion time is comparable or even shorter than the desorption time. These values of binding energies are typical of np-ASW, which are supposed to resemble more closely the ISM ices, but also match bare grains (see Tab.~\ref{tab:binding}). In Fig.~\ref{fig:analysis1} (right) the grey area shows the minimum and the maximum conversion time and includes the measurements found in the literature (symbols). 

This simple analysis gives an insight on the role played by the different physical ingredients, and suggests that the o-p H$_2$ conversion on grains might be relevant for cores with $T_{gas} \leq 15$~K. This slightly depends on the uncertainties on the binding energies and the conversion time.

\begin{figure}[!ht]
	\centering
	\includegraphics[scale=0.38]{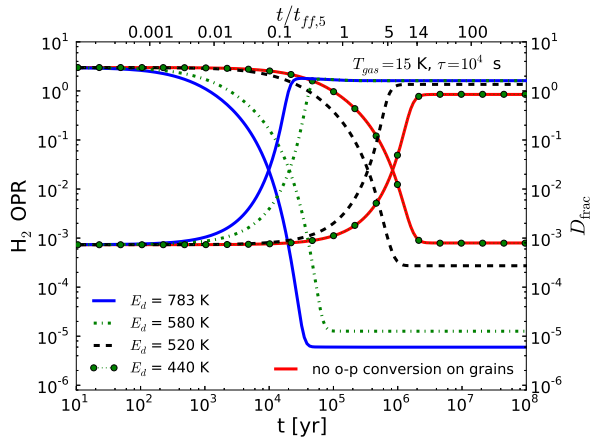}
	\caption{Effect of the binding energy on the evolution of OPR and $D_\mathrm{frac}$. The red solid line represents our reference run, i.e.~$T = 15$ K, $S = 1$, and no o-p H$_2$ conversion on grains. This is compared to models with o-p H$_2$ conversion on grains,  assuming $\tau_{conv} = 10^4$~s, but different binding energies, i.e.~different surface morphologies. The  free-fall time ($t_{ff,5}$) is evaluated at $n_\mathrm{H} = 10^5$~cm$^{-3}$.}\label{fig:analysis2}
\end{figure}

\subsection{One-zone model evolution}

\begin{figure*}
	\centering
	\includegraphics[scale=0.45]{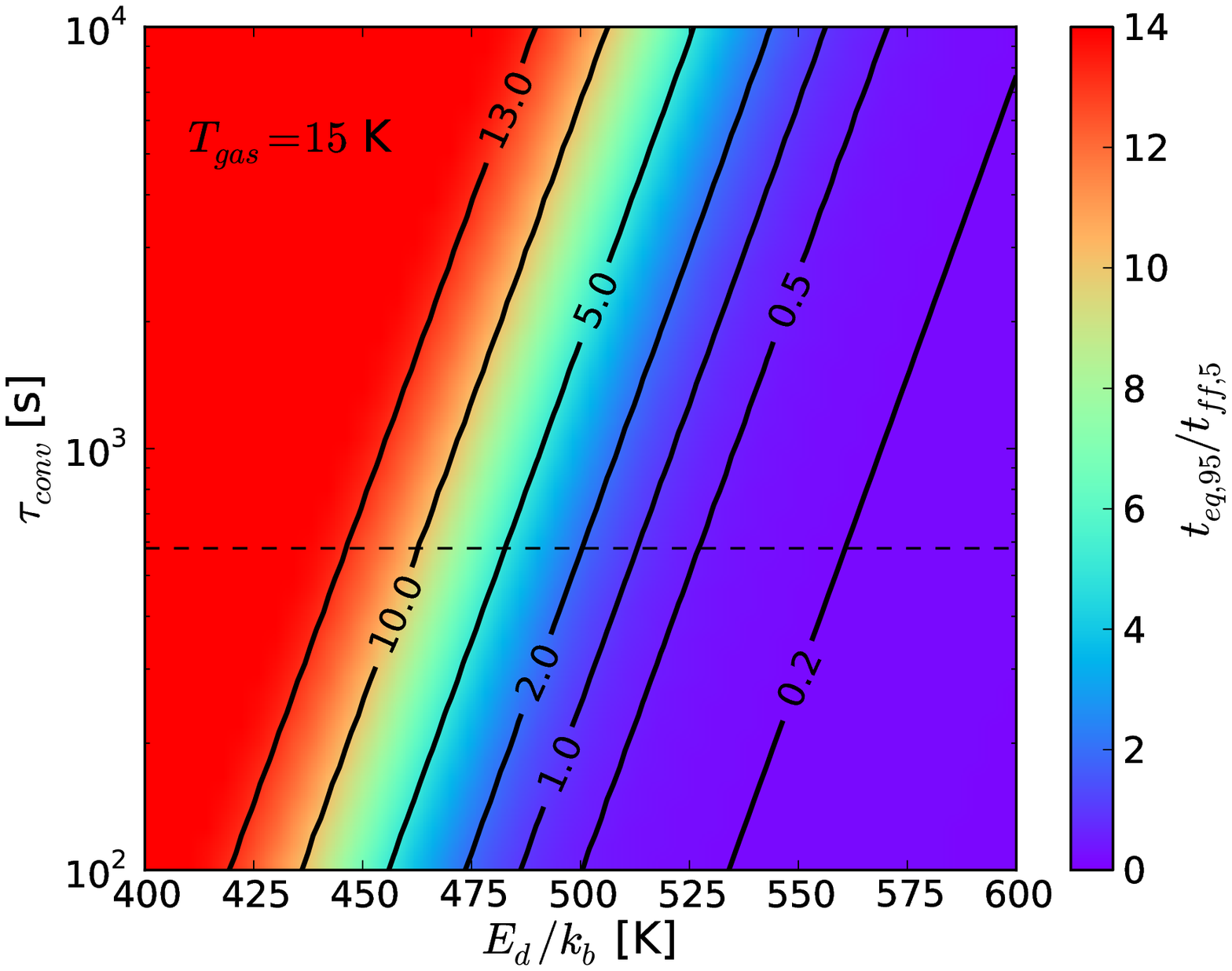}
	\includegraphics[scale=0.45]{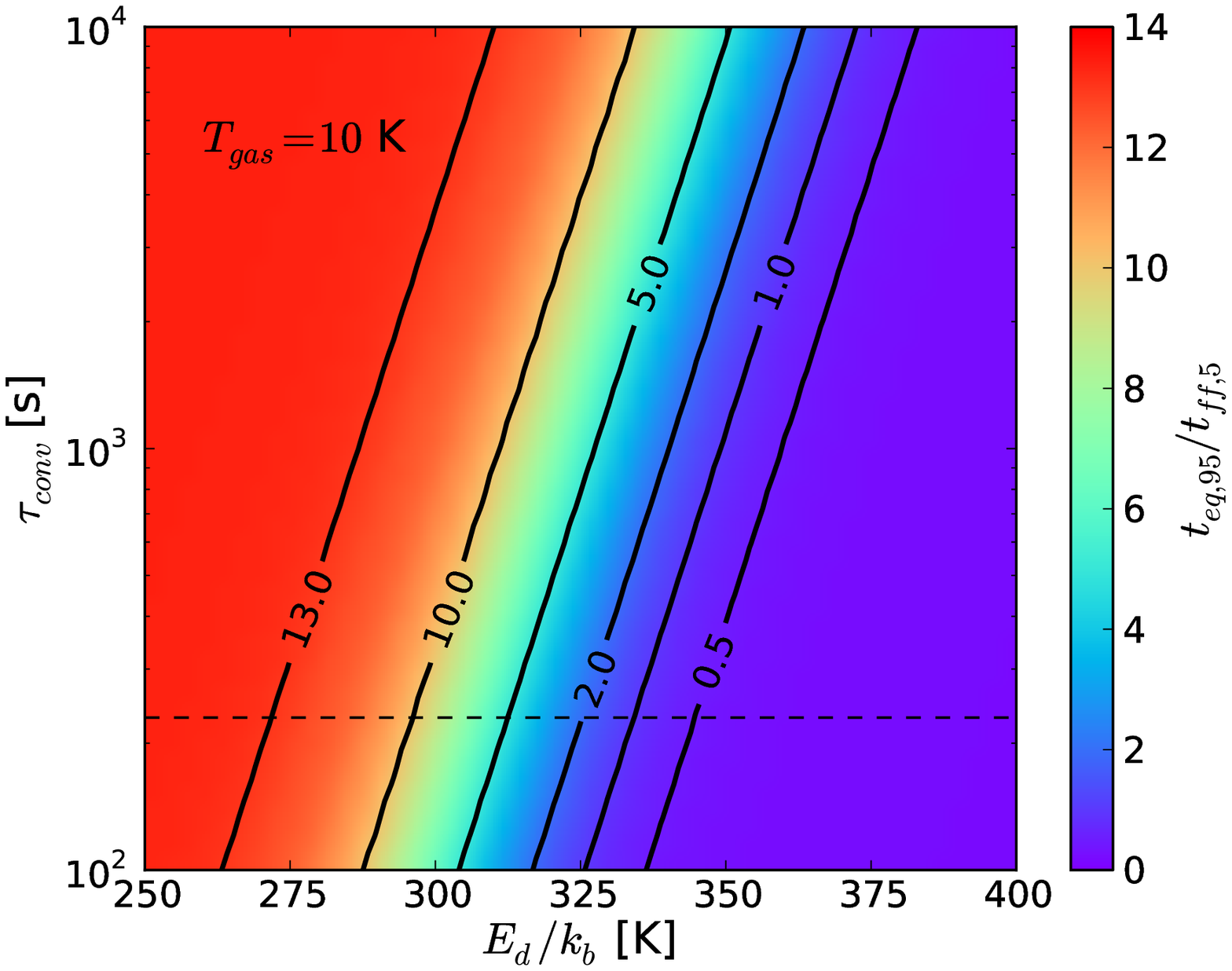}
	\caption{Effect of binding energy, conversion time, and temperature on the evolution of $D_\mathrm{frac}$ for a density of $n_\mathrm{H}=10^5$ cm$^{-3}$. The contours represent the ratio between the time needed to reach the 95\% of the equilibrium $D_{frac}$ value and the free-fall time. The dashed lines represent the shortest measured values of $\tau_{conv}$ at 10 and 15 K (right and left panel, respectively) as in \citet{Ueta2016} and \citet{Chehrouri2011}.}\label{fig:analysis3}
\end{figure*}

To confirm what we found from analytical considerations and give a more quantitative analysis, we present results from the time evolution of a one-zone model with \mbox{$n_{\rm H}$ = 10$^5$ cm$^{-3}$} and $T_{gas}=15$~K, which resembles conditions similar to the ones observed in molecular cloud cores
 \citep{Walmsley2004,Pagani2013,Kong2015}. Our initial conditions are $x_{\rm oH_2}$ = 3.75$\times 10^{-1}$, \mbox{$x_{\rm pH_2}$ =  1.25$\times 10^{-1}$}, and \mbox{$x_\mathrm{He}$ = 9$\times 10^{-2}$}, where $x_i = n_i/n_\mathrm{H}$, and we have assumed that the initial H$_2$ OPR is 3. 
Grains are initially charge-neutral with number density $n_{dust} = 6.38\times 10^{-10}  n_{\rm H}$. The assumed cosmic-ray ionization flux is $\zeta_{cr} = 2.5 \times 10^{-17}$~s$^{-1}$.

Fig.~\ref{fig:analysis2} shows the evolution of H$_2$ OPR and \mbox{$D_\mathrm{frac}$ = [H$_2$D$^+$]/[H$_3^+$]} for different binding energies by assuming the longest measured conversion time (\mbox{$\tau_{conv} = 10^4$ s}). Overall, the results reflect what we argued from analytic considerations on the timescales discussed in the previous Section. When the binding energy is above a critical value ($E_d \geq 520$~K),  the o-p H$_2$ conversion on dust starts to compete with the gas-phase conversion and has a strong impact already when $E_d=580$~K. In the latter case, the $D_\mathrm{frac}$ equilibrium at $15$~K is reached after $\sim 4 \times 10^4$~yr instead of $t \sim 4 \times 10^6$~yr, a difference of two orders of magnitude. This indicates that the timescale to reach high deuteration is shifted from 14~$t_{ff}$ to less than one $t_{ff}$, where $t_{ff}$ is the free-fall time. It is important to add that the o-p H$_2$ conversion on dust has also an impact on the H$_2$ OPR, which reaches lower equilibrium values in shorter times. As a consequence also the H$_2$D$^+$ OPR, which correlates very well with the H$_2$ OPR  \citep[see e.g.][]{Brunken2014} will follow a similar trend.

In Fig. \ref{fig:analysis3} we present a more extended analysis on the effect of the binding energy and conversion time on the time needed to reach equilibrium $D_{frac}$. We plot the ratio between $t_{ff}$ and the time needed to reach 95\% of the equilibrium $D_{frac}$, namely $t_{eq,95}$, at 10 and 15 K, respectively. The results can be explained in terms of three different regimes: i) an $E_d$ value below which no relevant o-p conversion takes place, ii) a transition binding energy range where an effect of a factor 2-5 is visible, and iii) the saturation regime, i.e. $E_d$ above which the o-p conversion on dust is always dominant with a change in $t_{eq,95}/t_{ff,5}$ of more than one order of magnitude. We note that the effect could be less strong when the initial density or the cosmic-ray flux are increased. 

These regimes are different depending on the temperature we consider: for a gas at 15 K the saturation regime starts at $E_d \sim 550$ K while at 10 K this threshold is much lower, $E_d \sim 370$ K. The latter is lowered to $E_d \sim 520$ K at 15 K and $E_d \sim 330$ K at 10 K when we assume the shortest measured conversion times (dashed line in the plots). We can conclude that in a range of temperatures of 10-15 K, and for binding energies above the discussed thresholds, o-p conversion on dust becomes relevant, with a weak dependence on the uncertainties affecting the dust physics. We note that the threshold $E_d(t_{eq,95} \equiv t_{ff})$ scales as $\propto - T_{gas} \ln(n_\mathrm{H})$. A shift towards larger (smaller) values is expected when we decrease (increase) the density or the temperature.

\begin{figure}[!ht]
	\centering
	\includegraphics[scale=0.38]{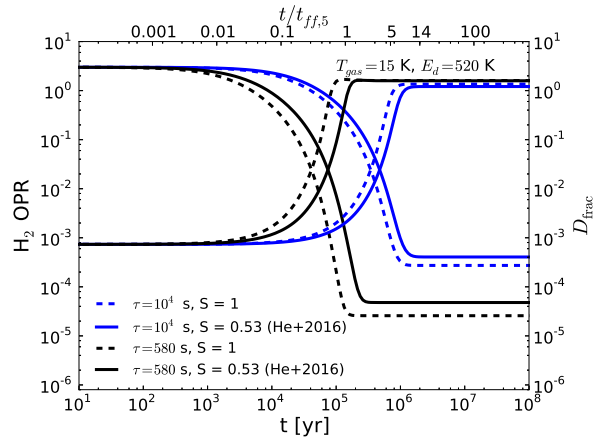}
	\caption{Run at 15~K for a binding energy of 520~K and two different conversion times. We compare runs at constant sticking coefficient $S=1$ and runs following \citet{He2016}.}\label{fig:analysis4}
\end{figure}

Finally, in Fig. \ref{fig:analysis4}, we show a set of runs where we employ a temperature-dependent sticking coefficient, obtained from laboratory measurements on np-ASW \citep{He2016}. The results are only slightly affected and are independent of the conversion time. 

Overall, our tests show that  \textit{binding energy and temperature}, i.e. the two quantities which affect the desorption time, \textit{are the most relevant, while the sticking coefficient and the o-p conversion time only slightly affects}  $D_\mathrm{frac}$.

\section{Discussion and Conclusions}
In this Letter we have explored the o-p H$_2$ conversion on grains, focusing on the imprint this process leaves on $D_\mathrm{frac}$, when it is more effective than the analogous process in the gas-phase. The H$_2$ o-p conversion on dust is regulated by different parameters, the most important being the binding energy and the conversion time. Great efforts have been made to provide measurements of these quantities with state-of-the-art laboratory experiments but not conclusive answers have been provided yet. 
Uncertainties due to the binding energy, coverage, and grain material, together with temperature strongly affect the outcome of chemistry on the surface. 
For the first time we have studied the impact of the o-p H$_2$ conversion on dust on the deuteration exploring the different uncertainties. From our analytic considerations and one-zone models we found that,  when above a given binding energy threshold, the uncertainties on the binding energies and conversion times, as well as the sticking coefficient, are marginally relevant. The binding energy determines the residence time of o-H$_2$ on grains, providing a time frame for the o-p conversion to occur. Our findings suggest a threshold value of binding energy in between $330-550$~K, when temperatures of $10 \leq T_{gas} \leq 15$ K and most recently measured conversion times are considered.  These values are in agreement with typical experiments on np-ASW but also with binding energies on bare silicates\footnote{No experiments on CO-rich ices (expected in cold dense cores) have been performed.}.

The effect of the temperature is in general more important. As the process is regulated by the desorption time (or the residence time), proportional to $\propto \exp(E_d/k_b T_{gas}$), an increase in temperature shortens the time available for the o-p conversion to occur. Increasing the binding energy or  decreasing the temperature, shift the timescale toward faster deuteration by two orders of magnitude in time, making the process more efficient. This provides an alternative route to explain the observed $D_\mathrm{frac}$ values and can affect the calculations of the cores chemical age. 

Despite the many uncertainties affecting laboratory experiments and their outcomes (i.e. binding energies, conversion times, sticking coefficient, and surface coverage), our results suggest that including the o-p H$_2$ conversion on grains in the modelling of deuteration in dense molecular cloud cores might play a crucial role when deuteration is employed as a chemical clock. We also stress that further laboratory measurements and theoretical calculations are needed to shed light on this important process, and that the implications should be explored in a fully dynamical simulation.

\acknowledgments
\section*{Acknowledgments}
We are grateful to Naoki Watanabe for useful discussion on the topic.
SB and DRGS thank for funding through the DFG priority program ``The Physics of the Interstellar Medium'' (projects BO 4113/1-2 and SCHL 1964/1-2). TG acknowledges the Centre for Star and Planet Formation funded by the Danish National Research Foundation. DRGS acknowledges funding via Fondecyt regular (project 1161247), the ``Concurso Proyectos Internacionales de Investigaci\'on, Convocatoria 2015'' (project PII20150171), ALMA--Conicyt (project 31160001) and the BASAL Centro de Astrof\'isica y Tecnolog\'ias Afines (CATA) PFB-06/2007. PC acknowledges the financial support of the European Research Council (ERC; project PALs 320620).

\end{document}